\documentstyle[11pt,newpasp,twoside,psfig]{article}
\def\vi{$V_{606}-I_{814}$\/ }
\def\vir{$V_{606}-I_{814}(r)$\/ }

\def\edcomment#1{\iffalse\marginpar{\raggedright\sl#1\/}\else\relax\fi}
\reversemarginpar

\begin{document}
\title{The Epoch of Galaxy Formation}
 \author{Raul Jimenez}
\affil{Dept of Physics \& Astronomy, Rutgers University, 136
Frelinghuysen Road, Piscataway NJ 08854. {\em raulj@physics.rutgers.edu}}

\begin{abstract}
I present a biased review of when the epoch of formation of galaxies
(both disks and ellipticals) maybe took place. I base my
arguments in simple (mostly) analytic models that have been recently
developed to reproduce most of the observed photometric, chemical and
dynamical properties of galaxies both at low and high redshift.
\end{abstract}

\section{Introduction}

A much sought after ``holy grail'' of cosmology is the epoch of
galaxy formation. Not only it will give us information about when the
intergalactic and intercluster medium was enriched (see this volume)
but also allows us to test (extremely) opposed scenarios of galaxy
formation and the nature of the initial conditions (e.g. Verde et
al. 2001).

The most direct method to compute the different predictions of galaxy
formation models consists in simulating in a computer the processes of
galaxy formation and evolution from different sets of initial
conditions. Unfortunately, the limitations in both computer power and
our knowledge of physics to simulate the dynamics of the gas and
somehow the dark matter itself, makes this approach inconclusive as to
what is the epoch of galaxy formation (see e.g. Ellis 2001). On the
other hand, great insight can be gained in understanding this epoch by
using simple analytic models of galaxy formation, that include robust
predictions from theory, and combining them with observations to
constrain the free parameters in the model. In the next two sections I
describe some progress using this hybrid approach in understanding the
epoch of galaxy formation for both disk and elliptical galaxies.

\section{Disk Galaxies}

Many authors have investigated galactosynthesis models for disk
galaxies, both locally and at high redshift, (e.g., Einseistein \&
Loeb 1996; Dalcanton, Spergel \& Summers 1997; Mo, Mao \& White 1998;
Jimenez et al. 1998; Avila--Reese et al. 1999; Somerville \& Primack
1999; van den Bosch 2000; Firmani \& Avila-Reese 2000; Mo \& Mao 2000;
Navarro \& Steinmetz 2000; Bullock et al. 2001; Boissier et al. 2001)
in which the properties of disk galaxies are determined primarily by
the mass, size, and angular momenta of the halos in which they form,
and which may contain the effects of supernova feedback, adiabatic
disk contraction, cooling, merging, and a variety of star-formation
(SF) recipes.  Additionally Buchalter, Jimenez \& Kamionkowski (2001),
investigated a variety of galactosynthesis models with realistic
stellar populations and made multi-wavelength predictions for the
Tully-Fisher (TF) relation.  With reasonable values for various
cosmological parameters, spin ($\lambda$) distributions,
formation-redshift ($z_f$) distributions, and no supernova feedback,
they produced an excellent fit to the local TF relation at all
investigated wavelengths ($B$, $R$, $I$, and $K$), as well as to
$B$-band TF data at $z=1$, and to the surface-brightness--magnitude
($\mu$-$M$) relation locally and at $z=0.4$.  These successes suggest
that their simplified mostly analytic approach captures the essential
phenomenology, even if it leaves out some of the details of more
sophisticated models. Using this model with a relatively minimal set
of ingredients it is possible to derive the most likely redshift of
formation of disk galaxies.

\begin{figure}[!tpb]
\centerline{
\psfig{figure=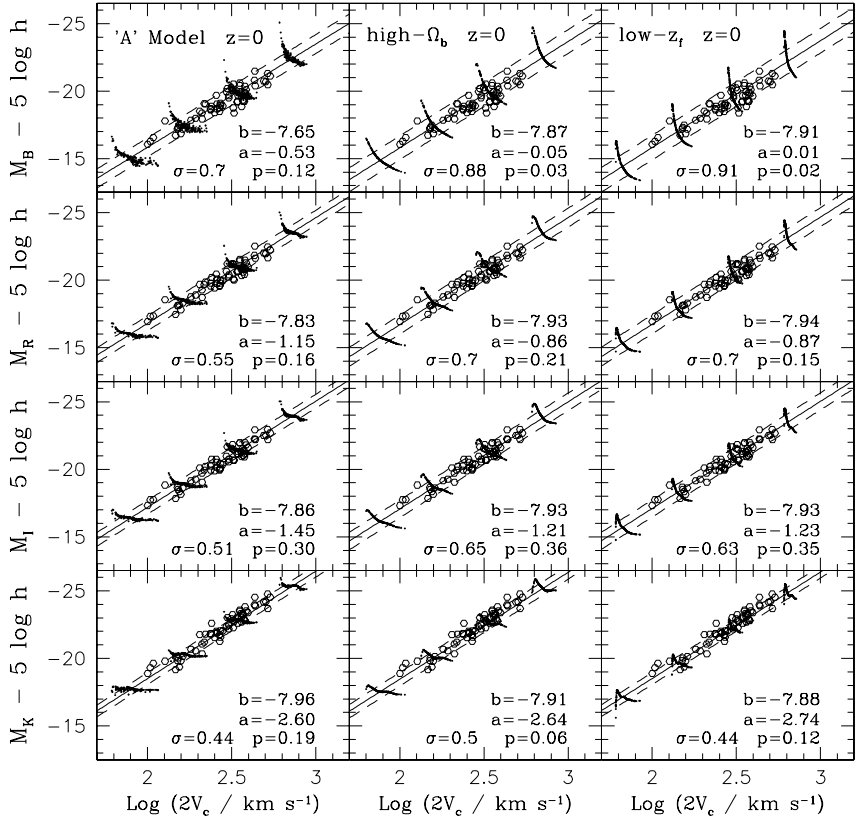,height=12cm,angle=0}}
\caption{Predicted $B$-, $R$-, $I$-, and $K$-band TF relations at
$z=0$ for the A model, high-$\Omega_b$ model, and low-$z_f$ model.
The scattered dots show the results for fixed masses
of $10^{10}$, $10^{11}$, $ 10^{12}$, and $10^{13}$ ${\cal M}_\odot$.  The
solid lines are the best-fitting TF relations, with zero points and
slopes given by $a$ and $b$, respectively, while the dashed lines show
the 1-$\sigma$ scatter, denoted in each plot by $\sigma$.}
\end{figure}

\subsection{Mostly analytic modelling of disk galaxies}

Here, I briefly review the main ingredients of the galactosynthesis
model developed in Buchalter, Jimenez \& Kamionkowski (2001), hereafter
BJK, based in turn upon that of Heavens \& Jimenez (1999). The authors
used the spherical-collapse model for halos (Mo, Mao \& White 1998),
the distribution of halo-formation times from the merger-tree
formalism (Lacey \& Cole 1994), and a joint distribution in $\lambda$
and $\nu$, the peak height (Heavens \& Peacock 1988). Halos were
treated as isothermal spheres with a fixed baryon fraction and
specific angular momentum, and their embedded gaseous disks, assumed
to form at virialization, have an exponential density
profile.\footnote{The assumptions of an isothermal profile and
effectively instantaneous disk formation constitute a great
oversimplification. BJK explore the impact of these assumptions and
conclude that, while severe, they do not bear a strong impact on the
predicted scaling relations explored here. The halo profile employed
in their work serves as an excellent approximation to a suite of
truncated-profile models everywhere except in the core. This
discrepancy, however, has little impact on the flat part of the
rotation curve with which we are concerned, or on the stellar
populations. The most significant effect is an uncertainty in the
normalization of the mass-circular velocity relation, but BJK find
that this primarily serves only to slide galaxies along the predicted
relations, resulting in little or no net change.  A complete
description of galaxy formation would of course require more detailed
modeling of the halo, and of halo-disk interactions.}  They implicitly
assumed that the halos of spiral galaxies form smoothly, rather than
from major mergers. An empirical Schmidt law relates the
star-formation rate (SFR) to the disk surface density (Kennicutt
1998).  A Salpeter initial mass function, a prescription for chemical
evolution, and a synthetic stellar-population code (Jimenez et
al. 1998) provide the photometric properties of disks at any $z$.

The model is defined by cosmological parameters and by the time when
the most massive progenitor contains a fraction $f$ of the present-day
mass, when a halo is defined to form.  They found that excellent
agreement with current data was obtained by their 'Model A', a
COBE-normalized $\Lambda$CDM cosmogony with $\Omega_0=0.3$, $h=0.65$,
$\Omega_b h^2 =0.019$, an untilted power spectrum with a shape
parameter given by $\Gamma=\Omega_0 h$, and with $f=0.5$.  The TF
relation in this model relied on both halo properties and upon the SF
history of the disk. The local TF scatter arose primarily from the
$z_f$ distribution, and secondarily from chemical evolution and the
$\nu$-$\lambda$ anticorrelation.  {\em In this model, disk formation
occurs primarily at $0.5 \la z \la 2$} and the TF slope steepens and
the zero points get fainter from $z=0$ to $z=1$.  Moreover, the amount
of gas expelled from or poured into a disk galaxy in this model is
relatively small and the disk and halo specific angular momenta are
equal.

The problem is that this fit is not unique and a suite of other models
that give good fits to the observations at low $z$ can also
be found.  To illustrate, they examined two alternative models, which
though less observationally favored, also meet the considerable burden
of yielding comparably good fits to the slope, zero-point, {\em and}
scatter of the TF relation at $z=0$ in $B$, $R$, $I$, and $K$. The
first is a CDM model with $\Omega_0=1$, $h=0.65$, constant metallicity
given by the solar value, $\lambda=0.05$ for all disks, an untilted
power-spectrum with amplitude $\sigma_8=0.5$ and empirical shape
parameter value of $\Gamma=0.2$, and $f=0.5$.  As shown in BJK and
elsewhere, high-$\Omega_0$ models generally produce disk galaxies too
faint to lie on the local TF relation.  To compensate for this, they
assumed $\Omega_b h^2 = 0.045$, and termed this the 'high-$\Omega_b$'
model.

Their second alternative is a $\Lambda$CDM cosmogony like Model A, but
with metallicity held constant at the solar value, $\lambda=0.05$ for
all disks, and $f=0.9$.  This results in a narrow distribution of
formation times peaked around $z\sim 0.2$ (for $L_*$-type disks), with
appreciable ongoing formation today and almost no halos forming
earlier than $z=1$. They thus denoted this the 'low-$z_f$' model.
This model will produce extremely young and bright
disks\footnote{Bulges, however, may be present at high $z$.}. To
compensate for this, they reduced the efficiency of SF in the
Schmidt-law prescription by 50\%.  This lower SF efficiency may be
plausible given the lower disk gas fractions predicted by their A
model as compared with observed values.

\begin{figure}[!tbp]
\centerline{
\psfig{figure=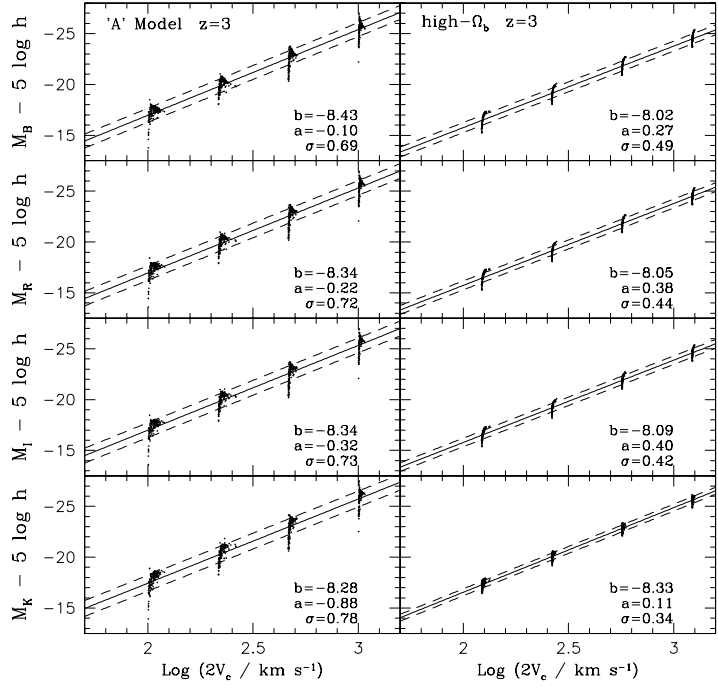,height=10cm,angle=0}}
\caption{TF predictions of the A model and high-$\Omega_b$ model at
$z=3$. The low-$z_f$ model predicts essentially no disks to have
formed at $z=3$ at these scales.}
\end{figure}

The left, middle, and right panels of Figure~1 respectively depict the
$z=0$ predictions for the A, high-$\Omega_b$, and low-$z_f$ models, in
the $B$, $R$, $I$, and $K$ bands.  In each panel, the solid line shows
a least-squares fit to the TF-relation prediction, with a zero-point
and slope given by '$a$' and '$b$', respectively, and 1-$\sigma$
errors given by the dashed lines and denoted in each plot by
'$\sigma$'. The four scattered-dot curves in each plot trace the
predicted TF spread for four fixed masses ($10^{10}$, $10^{11}$,
$10^{12}$, and $10^{13}M_{\odot}$), using $\sim 120$ points each.  The
open symbols represent extinction-corrected data from Tully et
al. (1998), comprised of spiral galaxies in the loose clusters of Ursa
Major and Pisces.  In each plot, the data are fit to the corresponding
model, and the value of $p$ gives the probability of obtaining a value
of $\chi^2$ as large as that measured, given that the model is
correct.  Since the data have excluded spirals that show evidence of
merger activity or disruption, they exclude from their predictions those
galaxies with $B-R < 0.3$.

Each of the three models yields a reasonable fit to the slope and
normalization of the TF relation in all wavebands.  Moreover, the
predicted scatter in the $B$, $R$, $I$, and $K$ bands roughly agrees
with the observed values of 0.50, 0.41, 0.40, and 0.41,
respectively. Yet the epoch of disk formation is rather different in
these three scenarios.

A way to break this degeneracy is to study the evolution of the TF
relation with redshift. The evolution of the TF relation, and in
particular its scatter, which owes to different mechanisms as one
looks at different wavebands and/or at different epochs, {\em can
probe the spread in halo $z_f$, as well as SF processes in the disk}.
Specifically, for local observations of evolved systems at redder
wavelengths, the scatter essentially decouples from the luminosity
axis since the light is tracing the total mass roughly independently
of the galaxy's age. By contrast, observations at high $z$ and/or in
bluer bands are more sensitive to the disk's age, resulting in a
scatter that couples more closely to the luminosity axis of the TF
relation, decoupling almost entirely from the $V_c$ axis in the case
of very young systems at $z=3$ (see Figure~2).

\section{Spheroidal Galaxies} 

Spheroidal galaxies are the main contributors to the enrichment of the
intergalactic medium. Therefore it is important to know when they
formed. Furthermore, field elliptical galaxies can be used to test
(extremely) opposed models of galaxy formation. The most obvious way
to discover the epoch of spheroid formation is to find the most
distant objects with spheroid morphology. This can be difficult for
two reasons: first, the morphology of a forming spheroid for the first
couple of Gyr may not be that of a spheroid, due to strong feedback
from recently formed stars and winds (e.g. Jimenez et al. 1999;
Pettini et al. 2001). Second, it might be extremely difficult to find
these progenitors due to their intrinsic low surface brightness;
e.g. a local elliptical galaxy becomes invisible for the HST at a
redshift larger than 2. Fortunately, some other indirect routes
exist. The most obvious one is to date the stellar population of the
spheroid galaxy from its composite stellar spectrum. This requires
very good $S/N$ spectrum and good stellar population models. The
dating of nearby ellipticals has been attempted recently by
concentrating on features that break the age--metallicity degeneracy
(e.g. Vazdekis \& Arimoto 1999; Trager et al. 2000) . This dating has
been done in the context of single stellar populations, i.e. it is
assumed that the elliptical galaxy is formed in a single burst of
infinitesimal duration at a given age for a single metallicity. Since
any tiny amount of star formation ($>5$\%) will dominate the optical
spectrum, this method is not optimal at determining the age of the
oldest stars in the elliptical galaxy. In fact, a wide range of ages
for local elliptical galaxies has been reported in these studies. A
more fruitful approach is to treat the star formation rate in
elliptical galaxies as a free parameter and try to do a
multi-parameter fit to the observed spectrum, this was attempted by
Reichardt, Jimenez \& Heavens (2001). In this work the authors found
that most elliptical galaxies in the small Kennicutt sample (Kennicutt
1992) had recent star formation activity albeit at a very low level
($\sim 5\%$), but the bulk of the population was formed at a redshift
higher than 2. Of course, the dating of the stellar population can be
done more successfully by looking at the most distant objects that
exhibit spheroid morphology at high--redshift, since then it is easier
to date the stellar population both because it is young and also
should be free of posteriori episodes of star formation. This has been
done recently by Dunlop et al. (1996), Spinrad et al. (1997), Dunlop
(1999) and Nolan et al. (2001) for two red spheroidal systems at
high--redshift. The ages of these objects yield a redshift of
formation, for this particular class of galaxies, larger than 4. A
different approach consists in using a simple but physically motivated
model to fit observational properties of elliptical galaxies in order
to determine their formation redshift. I devote the rest of this
section to describe such a model developed in Menanteau, Jimenez \&
Matteucci (2001).

\begin{figure}[!tbp]
\centerline{
\psfig{figure=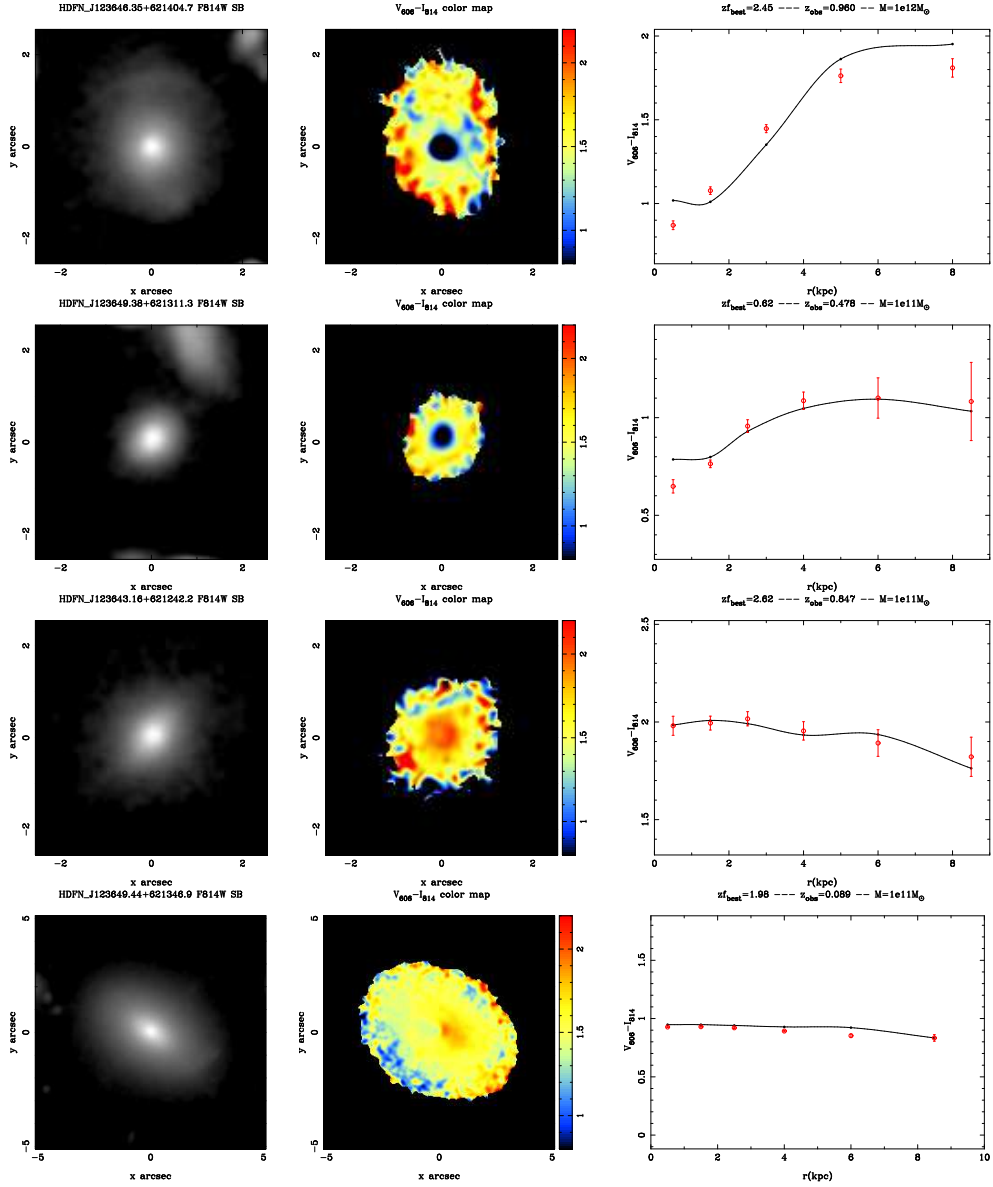,height=16cm,angle=0}}
\caption{An example of the ability of the model to reproduce a range
of different color gradients. Figure shows from left to right:
$I_{814}$--band surface brightness map, \vi color pixel map and \vir
color gradient. Open circle represent observed gradients while solid
line model prediction.}
\end{figure}

\subsection{Multi--zone modelling of elliptical galaxies}

In a recent paper (Menanteau, Jimenez \& Matteucci 2001), elliptical
galaxies were modelled as a system with spherical symmetry and
multiple zones. In particular it is assumed that the bulk ($> 80\%$)
of the gas in this model was in place at the time of formation and was
able to form stars, i.e.  was cool enough. The galaxy is then divided
in spherical shells --100 for the present case-- each of them
independent, i.e. no transfer of gas is allowed among shells. In each
of these shells star formation proceeds according to a Schmidt law:
SFR$=\nu \rho_{\rm gas}(t)$, where $\rho_{\rm gas}(t)$ is the volume
gas density in the shell and $\nu=8.6(M_{\rm
gas}/10^{12}M_{\odot})^{-0.115}$Gyr$^{-1}$. The initial mass function
is assumed to be a power law ($\phi(m) \propto m^{-0.95}$). Star
formation proceeds in each shell until the gas is heated up by SN to a
temperature $T$ that corresponds to the escape velocity of {\it each}
shell (see Martinelli, Matteucci \& Colafrancesco (2000) for a
detailed description of the model). The gas in the elliptical galaxy
is assumed to be within a dark matter halo of mass 7 times larger than
the gas mass ($\Omega_m=0.35$ and $\Omega_b=0.05$). The dark matter
follows the density profile described in Martinelli, Matteucci \&
Colafrancesco (2000). The chemical enrichment of the gas and stars is
followed in detail using up--to--date nucleosynthesis prescriptions
and taking into account the stars lifetimes. For each shell it is
assumed that mixing of the gas is very efficient in the whole shell
and shorter than the lifetime of the most massive stars.

For different masses the model predicts a different time dependence of
the SFR. In more massive systems the potential well will be deeper and
therefore it will take longer for the gas to reach temperatures larger
than the escape velocity in the potential, thus star formation will
last longer than in less massive systems. Also, for a fixed mass,
since the potential is deeper in the core of the galaxy, the model
predicts that star formation will last longer in the core than in the
outer regions (see Martinelli, Matteucci \& Colafrancesco (2000)). In
fact, the predicted SFR for this model is very similar to that from
detailed 1-D hydrodynamical models (e.g. Jimenez et al. (1999)). Knowing
the star formation rate and chemical composition for each radii, it is
then possible to compute the spectral properties of the galaxy by
using a synthetic stellar population code (Jimenez et al. 1998).

\subsection{Results}

The aim of Menanteau, Jimenez \& Matteucci (2001) analysis was to
directly compare the observed \vir color gradients with the
predictions of the above model.  Once the stellar mass of the galaxy
is determined from matching the $I-$band observations, the only free
parameter now in their model is the redshift of formation. To
accomplish this they used a maximum likelihood method ($\chi^2$) to
compute the most-likely redshift of formation ($z_{\rm F}^{\rm best}$)
for the best-fitting model. In order to avoid spurious results from
small fluctuations in the gradients due to noise, the model and data
gradients were re-binned to a common grid of $5-6$ shells up to a
maximum physical radius of $10$~kpc. In order to transform the
observed color gradients to physical (kpc) length, they assumed a flat
cosmology with $\Omega_m=0.35$ and $H_0=65$ km s$^{-1}$ Mpc$^{-1}$.

They applied this methodology to the whole sample of 77 E/S0 galaxies
from the HDFs. Figure~3 shows a selection of four representative
galaxies in the sample. It can be seen how the model can successfully
account for the observed range of \vir color gradients. It is worth
noting the ability of the model to reproduce the blue cores (inverse
steep gradients) present in the sample (Figure~3 upper two panels),
both the color difference and the physical scale at which this
occurs. In addition very flat and smooth color profiles can be
successfully reproduced (Figure~3 lower two panels) as expected from a
stellar population formed in single burst at high redshift.

Since the redshift of formation was the only free parameter in their
analysis, using the HDFs it was possible to determine this parameter
within the context of the above model.  Figure~4 shows the
distribution of the formation redshift for the whole sample. The top
axis shows the look--back--time. About 25\% of field E/S0 in their
sample have formed at $z > 4$, with $\sim 30\%$ of the sample having
formed at $z<1$. The medium redshift of formation is $z \sim 2$ and
therefore the medium age of the field ellipticals in the HDFs is $\sim
11$ Gyr. Therefore, as a whole, field ellipticals are only 1--2 Gyr
younger than cluster ellipticals, in agreement with the findings by
Bernardi et al. (1998) who compared the Mg$_2-\sigma_0$ relation for
field and cluster ellipticals. The main feature of fig.~4 is the
continuous formation of field E/S0 with redshift, they did not find
evidence for a {\it single} epoch of spheroid formation.  This result
should not be over interpreted though, due to the fact that the HDFs
cover a very small area of the sky and a large number of highly
clustered red objects have been found in larger area surveys (Cimatti
et al. 1999; McCarthy et al. 2000). It will be interesting to see what
color gradients these galaxies have and check weather or not inverse
color gradients are a common feature among elliptical galaxies at
high redshift.

\begin{figure}
\centerline{
\psfig{figure=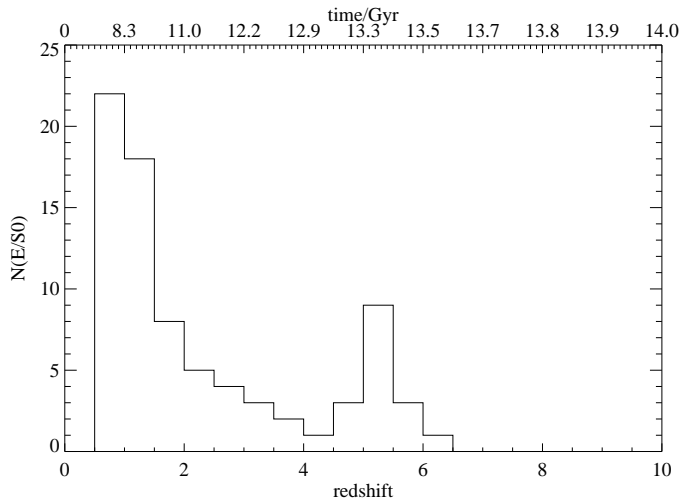,height=8cm,angle=0}}
\caption{Histogram for the most likely redshift of formation, $z_F$
values for the whole sample of E/S0.}
\end{figure}

\section{Conclusions}
I have argued that using simple (mostly) analytic models of galaxy
formation it is possible to understand the epoch of galaxy
formation. The Tully-Fischer relation gives us strong constraints as
when the big disks we see today can be assembled. Using this argument
the most likely redshift of formation for present disks is at $0.5 \la
z \la 2$. On the other hand, it has become apparent in recent years
that elliptical galaxies are not the simple systems once were thought
to be, i.e. single burst systems formed at high redshift. In fact, it
has been shown that some fraction of elliptical galaxies show
evidence of current star formation. Using a multi-zone model to
describe the formation of elliptical galaxies and that successfully
fits the color (and metallicity) gradients observed for spheroids in
the HDF, I have argued that there is not a single epoch for the
formation of spheroids and that it is better described as a continuous
process, although about 50\% of the spheroids seem to have formed at a
very high redshift ($z > 4$). Indeed, new telescopes (like SIRTF and
NGST) will shed new light on the epoch of galaxy formation issue.

\acknowledgements 
It is a pleasure to acknowledge my collaborators in this work: Ari Buchalter,
Marc Kamionkowski, Francesca Matteucci and Felipe Menanteau.

\end{document}